\documentclass[pre,floatfix,showpacs,showkeys,twocolumn,superscriptaddress,amsmath,amssymb]{revtex4-1}

\usepackage{bm}
\usepackage{soul}

\usepackage{graphicx}
\usepackage{color}

\newcommand{\grad}{\bm{\nabla}}
\newcommand{\vct}[1]{\mathbf{#1}}
\newcommand{\uvct}[1]{\mathbf{\hat{#1}}}
\newcommand{\maF}{\mathcal{F}}

\newcommand{\maH}{\mathcal{H}}

\newcommand{\maO}{\mathcal{O}}
\newcommand{\nabx}{\partial_{x}}
\newcommand{\naby}{\partial_{y}}
\newcommand{\theq}{\theta_{\mathrm{eq}}}
\newcommand{\aleq}{\alpha_{\mathrm{eq}}}
\newcommand{\Vex}{V_{\mathrm{ex}}}
\newcommand{\hpw}{h_{w}}
\newcommand{\hpweq}{h_{w}^{(\text{eq})}}

\begin{document}
\title{Stability of thin liquid films and sessile droplets under confinement}

\author{Fabian D{\"o}rfler}
\affiliation{Max-Planck-Institut f{\"u}r Intelligente Systeme,
Heisenbergstr. 3, 70569 Stuttgart, Germany} \affiliation{IV. Institut
f{\"u}r Theoretische Physik, Universit{\"a}t Stuttgart,
Pfaffenwaldring 57, 70569 Stuttgart, Germany}
\author{Markus Rauscher}
\email{rauscher@is.mpg.de}
\affiliation{Max-Planck-Institut f{\"u}r Intelligente Systeme,
Heisenbergstr. 3, 70569 Stuttgart, Germany} \affiliation{IV. Institut
f{\"u}r Theoretische Physik, Universit{\"a}t Stuttgart,
Pfaffenwaldring 57, 70569 Stuttgart, Germany} 
\author{S. Dietrich}
\email{dietrich@is.mpg.de}
\affiliation{Max-Planck-Institut f{\"u}r Intelligente Systeme,
Heisenbergstr. 3, 70569 Stuttgart, Germany} \affiliation{IV. Institut
f{\"u}r Theoretische Physik, Universit{\"a}t Stuttgart,
Pfaffenwaldring 57, 70569 Stuttgart, Germany}

\date{\today}

\begin{abstract}
The stability of nonvolatile thin liquid films and of sessile
droplets is strongly
affected by finite size effects. We analyze their stability within
the framework of density functional theory using the sharp kink
approximation, i.e., on the basis of an effective interface
Hamiltonian. We show that finite size effects suppress
spinodal dewetting of films because it is driven by a
long-wavelength instability. Therefore nonvolatile films are
stable if the substrate area is too small.
Similarly, nonvolatile droplets connected
to a wetting film become unstable if the substrate area is too large.
This instability of a nonvolatile sessile droplet turns out to be
equivalent to the instability of a volatile drop which can
atttain
chemical equilibrium with its vapor.
\end{abstract}

\pacs{68.08.Bc	wetting,\\\phantom{PACS numbers:} 68.43.-h	chemisorption/physisorption:
adsorbates on surfaces,\\\phantom{PACS numbers:}  68.03.Cd	surface tension and related
phenomena, \\\phantom{PACS numbers:} 82.60.Nh	thermodynamics of nucleation}
\keywords{wetting, nanofluidics}
\maketitle

\section{Introduction}

Dewetting of fluid films and the ensuing formation of sessile droplets are both
part of everyday experience. Moreover these mechanisms are
important for the
functioning of biological systems as well as for numerous
technological processes.
Dewetting on homogeneous substrates and the subsequent formation of droplets
have been studied both experimentally
\cite{reiter01a,seemann01b,seemann01d,becker03,muellerbuschbaum03,fetzer05,fetzer07b,hamieh07,ralston08} 
and theoretically
\cite{bausch94b,BeGrWi00,thiele01c,glasner03a,blossey06a,vilmin06b,bertrand07,deconinck08} in great detail.
Both mechanisms can be understood
quantitatively within the well established theory of wetting
phenomena \cite{degennes85,dietrich88,rauscher08a,rauscher10a}\/. More
recently, wetting and dewetting on structured surfaces 
receives
increasing attention, in particular with a view on controlling the dewetting
process on patterned surfaces 
\cite{jackman98,lipowsky00,kargupta01a,kargupta02a,brusch02,thiele03a,harkema03,checco12}
as well as in the context of microfluidics
\cite{dietrich05,squires05a,delamarche05,koplik06a,rauscher07a,mechkov08}\/.
Chemical patterns consisting of
lyophilic and lyophobic patches as well as topographic
patterns such as pits and grooves effectively lead to a lateral
confinement of wetting films and droplets.

It is well known that confinement modifies the structural and
thermodynamic properties of condensed matter. 
In small scale systems these finite size effects can either
stabilize or destabilize certain structures. For example, systems
exhibiting a long-wavelength instability are characterized by
a critical wavelength such that fluctuations with larger
wavelengths grow exponentially in time.
This type of instability is suppressed in systems smaller than
this critical wavelength. On the other hand, certain structures can
only exist if they are larger than a certain critical size, such as droplets
which, at least within classical nucleation theory, have to be
larger than the critical nucleus. This means that certain structures
are suppressed by finite size effects, or, to put it differently,
the availability of large space can stabilize them. 

Spinodally unstable flat films show a long-wavelength instability
such that the dependence of their stability on the substrate size is
obvious. Droplets of nonvolatile fluids, however, are
usually considered to be stable. But they are in chemical
equilibrium with an adsorbate or a wetting film connected to them
\cite{degennes85,dietrich88,rauscher08a,rauscher10a} which, on
very large substrates, acts like a reservoir: a spherical droplet of
$100$~nm radius has the same volume as an adsorbate layer 
with an effective thickness of 1~\AA\ on a substrate of $6.5\times
6.5\,\mu\mathrm{m}^2$\/. Therefore, an isolated droplet of a
\textit{nonvolatile} fluid placed on a
macroscopically large substrate is expected to be unstable with
respect to the formation of a film. 
If a droplet is \textit{volatile}, i.e., in chemical equilibrium with its
vapor, it is expected to be unstable, too, but with respect to
evaporation or condensation and the formation of an
equilibrium wetting layer.

The Laplace pressure in droplets decreases upon increasing their
diameter while the pressure in wetting films is determined by the
disjoining pressure. In the case of stable films it increases
with their thickness. In a stationary situation the pressure in the
droplet is balanced by the pressure in the connected film. Moving a small
amount of fluid from a droplet into its attached film increases
the pressure in the drop and, as the thickness of the film
increases, also the pressure in the film. However, due to volume
conservation, the larger the
substrate the smaller is the increase of the ensuing film
thickness and therefore the smaller is the increase of
pressure in the film. This implies, that beyond a certain substrate size
the pressure increase in the drop is larger than the
pressure increase in the film and the drop will dissipate into the
large film \cite{mechkov08}.

On the other hand, a substrate of limited size can
only support droplets with a base radius smaller than half the
substrate diameter. This means, that one can expect that there is a
window of droplet sizes for stable droplets as shown for
two-dimensional droplets with small slopes (i.e., liquid
ridges with a small contact angle) in Ref.~\cite{dutka12}\/.
Since droplet volumes scale with
the third power of the droplet radius while the volume of the
wetting or adsorbate film scales with the second power of the
substrate diameter, the influence of the wetting or adsorbate film on
droplet stability is
most important on the nanoscale because in this case the volumes of the
liquid in the droplet and in the film are comparable. In addition, due to the
non-vanishing
width of the three-phase-contact line there is a minimal
size for well defined droplets \cite{burschka93,bausch94b}, which
gives rise to an additional contribution to the finite size effects. 

In this spirit, here we study the influence of substrate size on the
stability of flat films and of droplets using the framework of
density functional theory within the sharp kink approximation, i.e.,
by minimizing the corresponding effective interface Hamiltonian
\cite{dietrich91b} in the presence of an effective interface
potential \cite{dietrich91a,napiorkowski92}\/.

\section{Effective interface Hamiltonian}

Within the capillary model for nonvolatile fluids
\cite{rowlinson02,safran03,brochardwyart04} interfaces and contact
lines are geometrical objects of zero volume and area,
respectively, and the free energy of a fluid in
contact with a substrate is given by bulk, interface, and line
contributions which are proportional to the volume, interface
areas, and contact line lengths, respectively. Within this
macroscopic model, finite size effects occur only if the
three-phase-contact line of a droplet reaches the lateral boundary of the
substrate. Wetting transitions and the dependence on temperature
and pressure of the thickness of wetting layers cannot be described
within this macroscopic model. 

For this reason, in order to access mesoscopic scales,
we resort to the effective interface model as 
the simplest non-trivial model to
describe a fluid in contact with a substrate. 
It can be derived from a classical 
density functional theory using the so-called sharp-kink approximation
\cite{dietrich91b,napiorkowski93}\/. As in the capillary model,
also in this approach interfaces are only two-dimensional mani\-folds
but contact lines, such as the three-phase-contact line between
fluid, vapor, and substrate have a nonzero width as a result of explicitly taking into
account the finite range of intermolecular interactions (for 
reviews see Refs.~\cite{rauscher08a,rauscher10a})\/. 
Accordingly, within this model line tensions emerge and are not
input parameters
\cite{indekeu92,dobbs93,indekeu94a,getta98,bauer99b,bauer00a,schimmele07}\/.

The effective, local interface
Hamiltonian $\maH$ for a liquid film in Monge parameterization
$z=h(x,y)$ on a homogeneous substrate with the
substrate-liquid interface $A$ located in the $xy$-plane reads
\begin{multline}
   \maH[h] = \int_{A}dx\,dy\, 
   \bigg[
    \sigma\,\sqrt{1+ (\nabx h)^2+ (\naby h)^2} \\
	 +\,\phi(h) +\delta \mu\,h
   \bigg] ,
\label{hamilgk}
\end{multline}
with the liquid-gas interface tension $\sigma$\/. 
$\phi(z)$ is the effective interface potential
\cite{dietrich88,dietrich91a,napiorkowski92} and it describes the
effective interaction between the liquid-vapor interface and the
liquid-substrate interface.
The last term $\delta\mu=\Delta\rho\,\Delta\mu$ is the
product of the undersaturation $\Delta \mu =
\mu_{\mathrm{coexistence}}(T)-\mu$ at temperature $T$ and the
number density difference
$\Delta \rho = \rho_\mathrm{liquid}-\rho_\mathrm{vapor}$ between
the coexisting phases, and thus it  measures the
thermodynamic distance from the bulk two-phase coexistence line.
Within mean field theory the
equilibrium configuration of the liquid-vapor interface
minimizes $\maH[h]$\/.

The Monge parameterization is restricted to single valued interface
configurations $z=h(x,y)$ so that droplets with contact angles larger than
$90^\circ$ cannot be described this way. Therefore we rewrite
Eq.~\eqref{hamilgk} in a parameter free form also used in the
finite element code employed below. For arbitrary parameterizations of
the liquid-gas interface $\vct{r}(u,v)=(x(u,v),y(u,v),z(u,v))$
the area of the surface
element is $d\vct{A}=\frac{\partial\vct{r}}{\partial u}\times
\frac{\partial\vct{r}}{\partial v}\,du\,dv = G\,\uvct{n}\,du\,dv$
with the interface normal vector $\uvct{n}$ pointing into the gas
phase and $G=\left|\frac{\partial\vct{r}}{\partial u}\times
\frac{\partial\vct{r}}{\partial v}\right|$\/. In the Monge
parameterization this reduces to $G=\sqrt{1+|\grad
h(\vct{r})|^2}$, i.e., the first term in the square brackets in
Eq.~\eqref{hamilgk}\/. The effective interface Hamiltonian can be
written in terms of an integral over the liquid-vapor interface
$S$
\begin{equation}
\maH[\vct{r}] = \int_{S}d\vct{A}\cdot \left\{ \sigma\,\uvct{n}(u,v)
+\left[\phi(z(u,v))+ \delta\mu\,z(u,v) \right]\,\uvct{e}_z\right\} ,
\end{equation}
with $\uvct{e}_z$ as the normal vector of the substrate-liquid
interface pointing into the liquid phase, i.e., in $z$-direction.

The existence of a classical density functional has been 
proven for grand canonical ensembles \cite{evans79}\/.
Nonetheless the functional in Eq.~\eqref{hamilgk} has been used
successfully to describe also equilibrium shapes of nonvolatile fluids
(i.e., in the canonical ensemble) by fixing the liquid volume
$V$ via a Lagrange multiplier $p$\/. In this case, $\delta\mu$ is
not an independent parameter. It turns out, that upon
adding the constant term $\delta\mu\,V$
(which is independent of the droplet shape) 
to the functional in Eq.~\eqref{hamilgk}, 
$\delta\mu$ and $p$ multiply
the same terms such that $\delta\mu$ can be absorbed into
$p$\/.
It will turn out
later (see Eq.~\eqref{elg}) that $p$ is the pressure
difference between the liquid and the vapor, and for droplets 
one has $p>0$, given the 
choice of sign for the Lagrange multiplier
contribution as in Eq.~\eqref{hamil}\/. Since in a nonvolatile
system the liquid and the vapor are not in thermodynamic
equilibrium, the pressures do not have to be equal.
This leads to a variation principle for the equilibrium shape of
the liquid-vapor interface of nonvolatile fluids. The
equilibrium shape $\vct{r}^{(\text{eq})}(u,v)$ minimizes
the functional ($\int_S d\vct{A}\cdot\uvct{e}_z = A$)
\begin{multline}
   \maF[\vct{r}(u,v)] = \int_{S}d\vct{A}\cdot
   \bigg\{
    \sigma\,\uvct{n}(u,v)+
	 \\ \left[\phi(z(u,v)) 
	 -  p\,\left( z(u,v)-\frac{V}{A}\right)\right]\,\uvct{e}_z
   \bigg\}.
\label{hamil}
\end{multline}

\begin{figure}
\includegraphics[width=\linewidth]{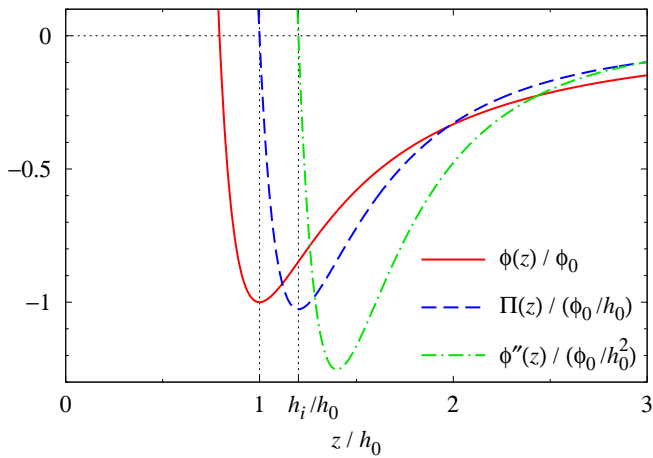}
\caption{The effective interface 
potential $\phi(z)$ (full red line) according to Eq.~\eqref{eip_model}
and the corresponding disjoining pressure $\Pi(z)=- \phi'(z)$
(dashed blue line) in units of $\phi_0$ and $\phi_0/h_0$,
respectively. The positions of the minimum of $\phi(z)$ at $z=h_0$
and of its \textit{i}nflection point at $z=h_i\equiv \sqrt[6]{3}\,h_0\approx 1.2\, h_0$ are indicated
by vertical dotted lines. Also shown is $\phi''(z)$ (dash-dotted
green line) which appears in the second variation operator
$\hat{O}_h$ in Eq.~\eqref{op} 
and which determines the stability of flat film solutions (see
Eq.~\eqref{crit2})\/.
\label{fig_eip_model}
}
\end{figure}

In the case of the laterally homogeneous substrates considered in
this paper, the effective
interface potential $\phi(z)$ does not explicitly depend on the
lateral coordinates $(x,y)$\/. However, due to the formation of
droplets one can still find non-trivial solutions to the
minimization problem in Eq.~\eqref{hamil}\/. The structure of $\phi(z)$
depends on the types of intermolecular interactions involved. The
simplest effective interface potential for 
long-ranged dispersion forces (described by Lennard-Jones type
interactions) and at temperatures below the wetting temperature
has the form
\begin{equation}
  \phi(z) = \phi_0 \,\left(\frac{h_0^8}{3\,z^8} -
  \frac{4\,h_0^2}{3\,z^2}\right).
\label{eip_model}
\end{equation}
The potential has a minimum of depth $-\phi_0$ at $z=h_0$ and an
inflection point at $z=h_i\equiv \sqrt[6]{3}\,h_0 \approx 1.2\,h_0$
(see Fig.~\ref{fig_eip_model})\/. The potential is negative for
$z>\sqrt[3]{1/2}\,h_0$ and approaches zero from below for $z\to
\infty$\/. The shape of $\phi(z)$ corresponds to that of a
continuous wetting transition \cite{dietrich88}.

\subsection{Minimizing the free energy functional}

Within mean-field theory the minimum of the 
effective interface functional $\maF$ containing the volume
constraint
(Eq.~\eqref{hamil}) renders the interfacial free energy for the corresponding 
stable equilibrium configuration. 

The functional in Eq.~\eqref{hamil} can be minimized numerically by 
means of an adaptive finite element algorithm implemented by the
software \textsc{Surface Evolver} \cite{brakke92}\/.
Therein, the liquid-vapor interface is represented by a
mesh of oriented triangles and, by means of a gradient projection
method, iteratively evolves towards the configuration of minimal
$\maF$ (for an example see Fig.~\ref{fig_nanodrops}) below\/.  

\subsection{Variations of the effective interface Hamiltonian}

Within the framework of variational calculus, a stable equilibrium profile 
corresponds to a vanishing
first variation and a negative second variation of the functional
$\maF$\/. In order to calculate them we return to the Monge
parameterization and introduce the perturbed interface
configuration $z=\tilde{h}(x,y)$ with
$\tilde{h}(x,y)=h(x,y)+\epsilon \,\Psi(x,y)$ and
$\tilde{p}=p+\epsilon\,\psi$, where $0<\epsilon\ll 1$ is a small
dimensionless parameter. It is straightforward
to show that the first variation $\delta^{(1)}\maF$ of
$\maF([\tilde{h}],\tilde{p}) = \maF([h],p) +
\epsilon\,\delta^{(1)}\maF + \epsilon^2\,\delta^{(2)}\maF +
\maO(\epsilon^3)$ with
respect to the interface configuration is given by ($h=h(x,y)$)
\begin{multline}
\label{delta1}
\delta^{(1)}\maF = \int_{A} dx\,dy\,\Psi\,\left[
-2\,\sigma \,H_h + \phi'(h)- p\right]\\
+\psi\,\int_{A}dx\,dy\,\left(h-\frac{V}{A}\right),
\end{multline}
with the mean curvature 
\begin{widetext}
\begin{equation}
\label{eqcurvature}
H_h = \frac{(\nabx^2 h)\,[1+(\naby)^2] -2\,(\nabx
h)\,(\naby h)\,(\nabx\naby h) + (\naby^2
h)\,[1+(\nabx h)^2]}{2\,\sqrt{1+(\nabx h)^2+(\naby h)^2}^3}
\end{equation}
\end{widetext}
of the unperturbed surface 
and $\phi'(h)$ denoting the derivative of the effective interface
potential with respect to the local film thickness.
The Euler-Lagrange equation corresponding to the vanishing of
$\delta^{(1)}\maF$ is
\begin{equation}
\label{elg}
2\,\sigma \,H_h + \Pi(h)+ p = 0
\end{equation}
together with
\begin{equation}
\label{elgvolume}
V = \int_{A} dx\,dy\,h ,
\end{equation}
where $\Pi(h)=-\phi'(h)$ is the disjoining pressure,
which describes the effective interaction between the
substrate surface and the film surface, and $2\,\sigma
\,H_h$ is the Laplace pressure, which follows from the
interface tension of the fluid surface. For equilibrium interface
configurations the sum of the disjoining pressure and of the Laplace
pressure is constant. The variation with respect to the
Lagrange multiplier $ p$ leads to the volume constraint (see
Eq.~\eqref{elgvolume})\/.

The second variation $\delta^{(2)}\maF$ of $\maF$ with respect to
the film height can be written as a form quadratic in
the perturbation $\Psi$:
\begin{equation}
\label{delta2}
\delta^{(2)}\maF =
\int_{A}dx\,dy\,\left(\Psi\,\hat{O}_{h}\,\Psi +
2\,\psi\,\Psi\right) ,
\end{equation}
with the self-adjoined operator
\begin{widetext}
\begin{equation}
\hat{O}_{h} = -\sigma \,
\left(\begin{array}{c} \nabx \\ \naby\end{array}\right)\cdot
\left(\begin{array}{cc} 
\frac{1+(\naby h)^2}{\left[1+(\nabx h)^2+(\naby h)^2\right]^{\frac{3}{2}}}
& 
\frac{(\nabx h)\,(\naby h)}{\left[1+(\nabx h)^2+(\naby h)^2\right]^{\frac{3}{2}}}
\\
\frac{(\nabx h)\,(\naby h)}{\left[1+(\nabx h)^2+(\naby h)^2\right]^{\frac{3}{2}}}
&  
\frac{1+(\nabx h)^2}{\left[1+(\nabx h)^2+(\naby h)^2\right]^{\frac{3}{2}}}
\end{array}\right) \cdot
\left(\begin{array}{c} \nabx \\ \naby\end{array}\right)
+ \phi''(h),
\label{op}
\end{equation}
\end{widetext}
and with the second derivative $\phi''(h)$ of the effective
interface potential. For the model potential given in Eq.~\eqref{eip_model}
$\phi''(h)$ is shown in Fig.~\ref{fig_eip_model}\/. It is positive
for small $h$ and negative for large $h$\/.
The second variation with respect to the Lagrange
multiplier is identical to zero. The mixed variation 
with respect to $p$ and $h$ leads to the second term in
Eq.~\eqref{delta2} which due to $\psi = \textit{const.}$ vanishes
for perturbations $\Psi(x,y)$ which conserve the volume.
The stability of a solution of the Euler-Lagrange
equation \eqref{elg} is determined by the spectrum of eigenvalues 
of $\hat{O}_{h}$\/. A solution is linearly stable if all eigenvalues
are positive. 

\section{Thin films and nano-droplets}

On a chemically homogeneous substrate with an area $A$ there
exist two distinct classes of solutions of the Euler-Lagrange
equation \eqref{elg}. One consists of \textit{f}lat films with
\begin{equation}
\label{def_hf}
h^{(\text{eq})}(x,y)=h_f=V/A.
\end{equation}
The other class consists of nontrivial droplet solutions with one or many
droplets smoothly connected to a wetting film. Here we
focus on solutions with a single droplet 
because in general two or more droplets connected via a
wetting film are unstable with respect to coarsening. 
In the following we discuss the stability of flat films and such
droplets as a function of the substrate area $A$, of the excess liquid
volume 
\begin{equation}
\label{defVex}
\Vex = V - A\,h_0=(h_f-h_0)\,A,
\end{equation}
and of material properties encoded in $\phi(h)/\sigma$\/.

\subsection{Flat films}

For flat films with homogeneous thickness $h_f$ the
Euler-Lagrange equation \eqref{elg} reduces to
\begin{equation}
    p + \Pi(h_f) = 0.
\label{elg_planar}
\end{equation}
This means that for any size of the substrate area a homogeneous
flat film obeying Eq.~\eqref{elg_planar}
is obviously a solution of the Euler-Lagrange equation. It
represents either a local maximum, a local minimum, or a saddle
point of the free energy functional in Eq.~\eqref{hamil}\/.
The curvature of the interface is zero and thus the liquid gas interface tension
drops out. If $h_f$ minimizes the effective interface potential one
has $ p=0$ (assuming that $\phi(h)$ is differentiable)\/.
For a flat interface the operator $\hat{O}_{h}$ in Eq.~\eqref{op}, 
which determines the linear stability of the flat film solution,
reduces to
\begin{equation}
   \hat{O}_{h}= -\sigma \,(\nabx^2 + \naby^2)+\phi''(h_f).
\end{equation}
The corresponding eigenvalue problem has the form of a stationary single
particle Schr\"odinger equation with a potential which is constant
across the domain of the substrate. In Fig.~\ref{fig_eip_model}
$\phi''(z)$ is shown for the model potential from
Eq.~\eqref{eip_model}\/. The inverse surface tension plays the role of the
mass.

The eigenvalue spectrum of this operator depends on the shape of
the domain and on the boundary conditions at its borders. Boundary
conditions corresponding to actual substrates of finite size are
rarely compatible with a flat film solution because usually there is
a bending of the interface at the edge of the domain. For example,
at the edge of a lyophilic patch on a lyophobic substrate the
film thickness will go to zero (or at least to a microscopically
small value) and at the brim of a flat piece of substrate the fluid
film either continues onto the side walls or ends with thickness
zero. The two simplest types of mathematical boundary conditions,
which allow for flat film solutions, are either periodic boundary
conditions or a Neumann type boundary condition which corresponds
to zero slope of the film surface at the domain boundaries. The
latter would correspond to upright side walls with an equilibrium
wetting angle of $90^\circ$ at a pit-shaped substrate. However,
even for such a setup, the interplay of the long-ranged forces from the
substrate and from the side wall would lead to a bending of the
film surface \cite{moosavi06b,moosavi09}\/.

For a square substrate with edge length $L=\sqrt{A}$ and with
Neumann boundary conditions the eigenvalue problem
corresponding to $\hat{O}_{h}$ can be factorized by separating
the variables and the eigenfunctions are
given by plane waves. The degeneracy of the eigenfunctions
characterized by
wave vectors of equal modulus is alleviated by the boundary
condition. Assuming the two edges
of the substrate to be aligned with the $x$-axis and with the
$y$-axis, respectively, the eigenfunctions are given by 
\begin{equation}
   \Psi_{nm}(x,y) \propto\
	\cos\left(\tfrac{2\,\pi\,n}{L}\,x\right)\,
	\cos\left(\tfrac{2\,\pi\,m}{L}\,y\right),
\end{equation}
with $n,m\in \mathbb{N}_0$. Since $\Psi_{(-n)m}=
\Psi_{n(-m)}=\Psi_{(-n)(-m)}=\Psi_{nm}$ we only consider non-negative
indices. Since we consider a nonvolatile
system there is volume conservation, i.e., 
$\int_{A}dx\,dy\,\Psi_{nm} =0$ and therefore either $n$ or
$m$ have to be positive. The corresponding eigenvalues are
given by
\begin{equation}
\lambda_{nm}=\sigma \,\left(\tfrac{2\,\pi}{L}\right)^2 \,\left(n^2 + m^2 \right) +\phi''(h_f)\/.
\end{equation}
Therefore the film is linearly stable, i.e.,
$\min\limits_{n,m}\lambda_{nm}>0$, for 
\begin{equation}
\label{crit2}
\frac{(2\,\pi)^2}{A}
>-\frac{\phi''(h_f)}{\sigma }.
\end{equation}
For substrates of infinite, i.e., macroscopic, size $A$ this is the case
only if $\phi''(h_f)>0$\/. For the model effective interface potential in
Eq.~\eqref{eip_model} the latter inequality holds for
\begin{equation}
h_f<h_i\equiv \sqrt[6]{3}\,h_0\approx
1.2\,h_0,
\label{crit1}
\end{equation}
Since $h_i>h_0$ (see Fig.~\ref{fig_eip_model}) films with negative
excess volumes (i.e., $h_f<h_0$ (see Eq.~\eqref{defVex})) exhibit
$\phi''(h_f)>0$ so that, according to  Eq.~\eqref{crit2}, they are 
linearly stable for any substrate size $A = L^2$\/.
However, even for $\phi''(h_f)<0$
flat films are linearly stable if the substrate size $L$ is
below the critical value $L_c=2\,\pi\,\sqrt{\sigma/|\phi''(h_f)|}$.
This perturbation
analysis does not yield any information about the nonlinear
stability of film solutions, i.e., whether a flat film has
a lower free energy than a droplet.

\subsection{The nano-droplet configuration}

\begin{figure}
\includegraphics[width=\linewidth]{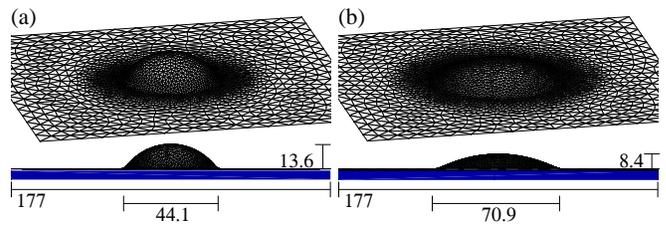}
\caption{ 
Interface configurations $h^{(\text{eq})}$ of nano-droplets as
obtained by numerical minimization of the functional $\maF$ in
Eq.~\eqref{hamil} based on the effective interface potential in
Eq.~\eqref{eip_model}
for $A/(\pi\,h_0^2)=100^2$ and under the constraint
$\Vex/(A\,h_0)=0.5$, with $\phi_0/\sigma =0.5$
($\theq=60^\circ$, left panel (a)), and $\phi_0/\sigma =0.1$
($\theq=26^\circ$, right panel (b))\/.
In the projected side view (bottom row), the underlying substrate of area $A$
is indicated in blue. Lengths are given in units of $h_0$\/.
Drop heights are measured from above the wetting film thickness
$\hpweq=1.009\,h_0$ in (a) and $\hpweq=1.012\,h_0$ in (b)\/.
During the iterative minimization process, the mesh size of
the triangulation has been coupled to the evolution of the
interface shape in an adaptive way in order to optimize the spatial
resolution locally.  The lateral boundary conditions are
implemented by a
constraint on the boundary vertices, such that their lateral
coordinates are fixed during the minimization process while the
perpendicular height coordinate can evolve freely, effectively
corresponding to neutral wetting (contact angle $90^\circ$) at
vertical side walls (not shown) or Neumann boundary conditions.
}
\label{fig_nanodrops}
\end{figure}

For a given area $A$ and a certain ratio
$\Vex/(A\,h_0)$ (see Eq.~\eqref{defVex}), nano-droplets with a nonzero 
pressure $ p>0$ minimize the free energy $\maF$ in
Eq.~\eqref{hamil}\/.
This is due to the interplay of the surface free energy
densities and the effective interface
potential, in combination with the non-volatility of the liquid and
the finite area $A$ of the solid-liquid interface.
Since the difference between the liquid-substrate
and the gas-substrate surface tensions is given by 
$\sigma +\phi(h_0)$ Young's law \cite{young05} reads
\cite{dietrich88}:
\begin{equation}
\cos \theq = 1-\frac{\phi_0}{\sigma }.
\label{young}
\end{equation}
$\theq$ denotes the equilibrium contact angle of a macroscopic
drop.
The influence
of the ratio $\phi_0/\sigma $ on the shape of a nano-droplet is shown in 
Fig.~\ref{fig_nanodrops}\/. 
A suitable definition of the contact angle of a nano-droplet is to
determine the curvature of its surface at the apex, to inscribe the corresponding
cap of a sphere which intersects the asymptote of the attached
wetting film thus forming a contact angle \cite{schimmele07}\/. 
For the systems studied here, this contact angle is smaller than
$\theq$\/.

The wetting
film surrounding the nano-droplet is almost flat, i.e.,
$2\,\sigma\,|H_h| \ll |\Pi(h)|$ (see Eq.~\eqref{elg}). According
to this
Euler-Lagrange equation \eqref{elg}, the
spatially constant pressure 
$ p$ is approximately given by
\begin{equation}
    p \approx -\Pi(\hpweq),
\label{elg_pw}
\end{equation}
and thus $\hpweq>h_0$ implies $ p >0$
(see Fig.~\ref{fig_eip_model})\/.

The height $\hpweq$ of the wetting film, the
pressure $ p$, the disjoining pressure
$\Pi(\hpweq)$ of the wetting film,
and the ratio between the drop free
energy $\maF_\text{drop}$ and the free energy $\maF_\text{film}$ 
of a flat film with the same excess volume are shown in Table~\ref{table} 
for several values of $\Vex$.
For decreasing values of $\Vex/(A\,h_0)$ with constant $A$,
$ p$ increases. This is mainly due to the increasing
curvature of the liquid-vapor interface. For the same reason the
pressure in macroscopic drops
also increases with decreasing volume. While the free energy
$\maF_\text{drop}$ of large drops turns out to be smaller than the free energy
$\maF_\text{film}$ of a
flat film with the same excess volume, the
situation is reversed for $\Vex/(A\,h_0) < 0.06$ (the critical
excess volume lies between $0.05\,A\,h_0$ and $0.06\,A\,h_0$)\/. This
means that nano-droplets below a certain size 
become metastable or unstable.

\begin{table}
\begin{center}
\begin{tabular}{|c|c|c|c|c|}
\hline
$\Vex/(A\,h_0)$ & $\hpweq/h_0-1$ & $
p\,h_0/\sigma $ & 
$-\Pi(\hpweq)\,h_0/\sigma $ &
$\maF_\text{drop}/\maF_\text{film}$ \\
\hline
0.05 & 0.0263 & 0.1777 & 0.1779 & 1.0003 \\ 
0.06 & 0.0219 & 0.1526 & 0.1522 & 0.9975 \\ 
0.10 & 0.0166 & 0.1191 & 0.1192 & 0.9782 \\ 
0.20 & 0.0125 & 0.0922 & 0.0922 & 0.9130 \\ 
0.50 & 0.0091 & 0.0674 & 0.0684 & 0.7730 \\ 
\hline
\end{tabular}
\end{center}
\caption{
Characteristics of nano-droplets obtained via
numerical minimization of $\maF$ in Eq.~\eqref{hamil} based on
Eq.~\eqref{eip_model} for
$A/(\pi\,h_0^2)=100^2$ and $\phi_0/\sigma =0.5$;
$\hpweq$ is the height of the film at the edge of
the numerical domain.
The pressure $p$ as the value of the Lagrange
multiplier for fixing the volume and the disjoining pressure 
$\Pi(\hpweq)$ of the wetting film surrounding the
droplet as calculated from the numerically determined $\hpweq$
are balanced according to Eq.~\eqref{elg_pw}\/. Accordingly,
the
differences between the third and fourth column 
indicate the level of numerical accuracy.
$\maF_\text{drop}/\maF_\text{film}$ is the ratio of the
(mean-field)
surface free energy of a nano-droplet and the free
energy of a flat film with a height
$h_f=\Vex/A+h_0$\/.
}
\label{table}
\end{table}

\subsection{Morphological transition}

\begin{figure}
\includegraphics[width=\linewidth]{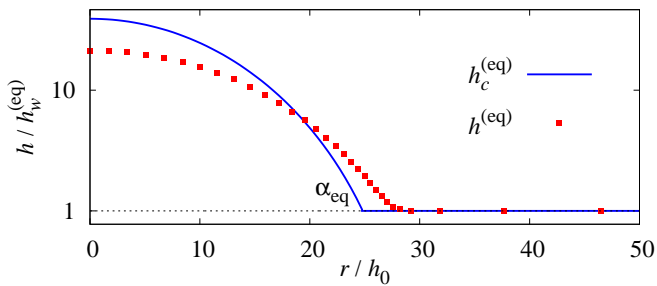}
\caption{
Vertical cut through the apex of a fully numerically obtained interface profile
$h^{(\text{eq})}$ (red squares)
and the corresponding approximate
profile $h^{(\text{eq})}_c$ (full blue line) consisting of a spherical cap
resting on a flat film. The profiles correspond to $A/(\pi\,h_0^2)=100^2$,
$\Vex/(A\,h_0)=0.5$, and $\phi_0/\sigma =
0.5$, which are the parameters corresponding to 
Fig.~\ref{fig_nanodrops}(a) and to the bottom line in
Table~\ref{table}\/. Although the domain for the numerical
calculation is
rectangular, the droplet shape is to a good approximation radially
symmetric ($r=\sqrt{x^2+y^2}$)\/.
The wetting film height for both profiles is
$\hpweq=1.01\,h_0$ and the contact angle is
$\aleq=55^\circ$ compared with $\theq= 60^\circ$ for the
corresponding macroscopic
drop. The free energies for these profiles
agree up to the third digit.
}
\label{fig_profiles}
\end{figure}

In order to analyze the morphological phase transition between
nano-droplets and flat films as indicated by the numerical data
discussed above, we minimize the effective interface Hamiltonian
$\maF$ in Eq.~\eqref{hamil} in the  subspace of
interface shapes $h_c(x,y)$ describing a spherical
\textit{c}ap sitting on top of a
flat wetting film (see Fig.~\ref{fig_profiles})\/. For a given
total volume of liquid, these trial profiles are parameterized by the contact
angle $\alpha$ and the wetting film height $\hpw$\/. The
latter determines the fluid volume available for the drop connected
to the film and the
contact angle determines the drop shape. This ansatz reduces the problem
of minimizing $\maF$ in Eq.~\eqref{hamil}  
to a minimization problem of the function 
\begin{equation}
F(\alpha,\hpw) =  \maF[h_c(x,y)] 
\label{conv_F}
\end{equation}
depending on the two variables $\alpha$ and $\hpw$ with the
minimum at $\aleq$ and $\hpweq$\/. The
corresponding minimizing profile is denoted by
$h_c^{(\text{eq})}$\/.
In contrast to the direct, full numerical minimization of the free energy
functional in Eq.~\eqref{hamil}, the function
$F(\alpha,\hpw)$ provides also a free energy landscape in the
parameter space $(\alpha, \hpw)$\/. Since for these
two-parameter trial functions the wetting film is perfectly flat,
the Laplace pressure vanishes and instead of
Eq.~\eqref{elg_pw} one has
\begin{equation}
\label{elg_pw2}
p=-\Pi(\hpw^{(eq)}).
\end{equation}

\begin{figure}
\includegraphics[width=\linewidth]{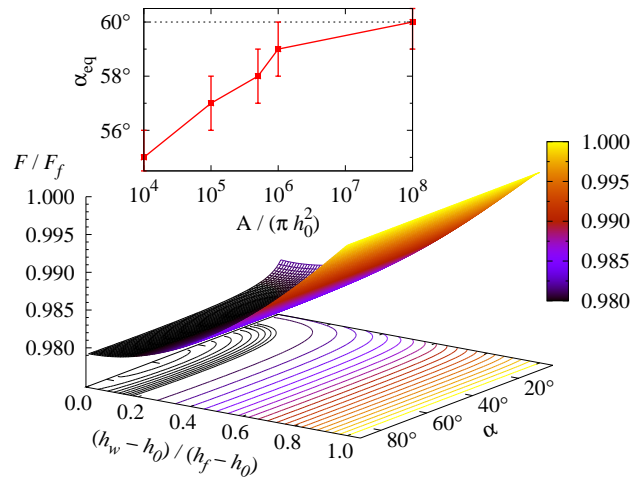}
\caption{
The approximate interfacial free energy $F(\alpha,\hpw)$
close to the macroscopic limit as described by
Eq.~\eqref{macro_limit}:
$A/(\pi\,h_0^2)=10^8$, $\Vex/(A\,h_0)=0.06$, and for
$\phi_0/\sigma =0.5$ corresponding to
$\theq=60^\circ$\/. 
$F_f=A\,(\sigma +\phi(h_f))$ is the free energy of
the flat film solution for these parameters; $h_f-h_0 = \Vex/A$\/.
The global minimum is located at $\alpha\approx
\theq=60^\circ$ and $\hpw\approx h_0$\/. 
The contour lines are almost parallel to the $\alpha$-axis. 
(The contour lines shown range from $0.9792$ to $0.98$ in steps of $0.0001$ and
from $0.98$ to $1.0$ in steps of $0.001$.)
The inset shows the equilibrium angle $\aleq$ upon approaching the macroscopic
limit as described by Eq.~\eqref{macro_limit} for the same excess
volume as used in the main
figure; $\aleq$
approaches $\theq$ from below. The error bars are due to
numerical inaccuracies. With $h_f/h_0 = \Vex/(A\,h_0)+1 =1.06
<h_i/h_0=1.2$ (see Eq.~\eqref{crit1}) the flat film solution with
$(\hpw-h_0)/(h_f-h_0)=1$ is expected to be linearly stable (i.e.,
metastable)\/. But the number of data points
calculated here is to small in order to be able to
detect the corresponding free energy
barrier (cf. Fig.~\ref{fig_as100} for a smaller substrate)\/.
\label{fig_Fmacro}}
\end{figure}

In the macroscopic limit, i.e., upon increasing both $A$ and
$\Vex$ such that 
\begin{equation}
   \frac{A}{h_0^2} \to \infty
   \quad\text{with}\quad
   \frac{\Vex}{A\,h_0} = \text{const}
\label{macro_limit}
\end{equation}
one finds $\hpweq\to h_0$ for the droplet solution because in this limit the
Laplace pressure $2\,\sigma\,H_h$ as well as the disjoining
pressure at the cap
apex vanish. The reason for this
is that the curvature of the droplet surface goes to zero if the
drop size diverges and that
the disjoining pressure vanishes for large distances from the
substrate surface.
Therefore the sum of the disjoining pressure and of the Laplace
pressure, i.e., $-p$, also vanishes (see Eq.~\eqref{elg})\/.
The Lagrange
multiplier $p$ does not depend on the position along the droplet
surface and, 
according to Eq.~\eqref{elg_pw2},
the disjoining
pressure on the wetting film is also zero. 
Therefore a macroscopic liquid cap with volume $\Vex$ is
formed above the level $\hpweq=h_0$ where $\Pi(h_0)=0$\/.
The numerical minimization of $F(\alpha,\hpw)$ also yields, in this
limit, $\aleq \to \theq$ with $\theq$ given by Eq.~\eqref{young}\/.
Figure~\ref{fig_Fmacro} shows the free energy landscape
$F(\alpha,\hpw)$ for a large drop.  All points in the parameter
space $\{\alpha, (\hpw-h_0)/(h_f-h_0)<1\}$ correspond to
droplet solutions (see
Eq.~\eqref{def_hf}), i.e., $\alpha\ne 0$\/. The line 
$(\hpw-h_0)/(h_f-h_0)=1$ corresponds to a flat film solution
for which $F(\alpha,\hpw)$ is independent of $\alpha$ because
the volume of the droplet is zero.
The global minimum of the free energy is located at $\aleq\approx
\theq=60^\circ$ and $\hpweq\approx h_0$\/. 
The contour lines of the free energy landscape 
close to the minimum in Fig.~\ref{fig_Fmacro} 
are almost parallel to the $\alpha$ axis and hence shape
fluctuations of the liquid cap with a constant cap volume are more
likely than volume fluctuations, i.e., fluctuations of the
wetting film height $\hpw$\/. As shown in the inset of
Fig.~\ref{fig_Fmacro} the equilibrium angle $\aleq$ approaches
the macroscopic equilibrium contact angle $\theq$ from below.

\begin{figure}
\includegraphics[width=\linewidth]{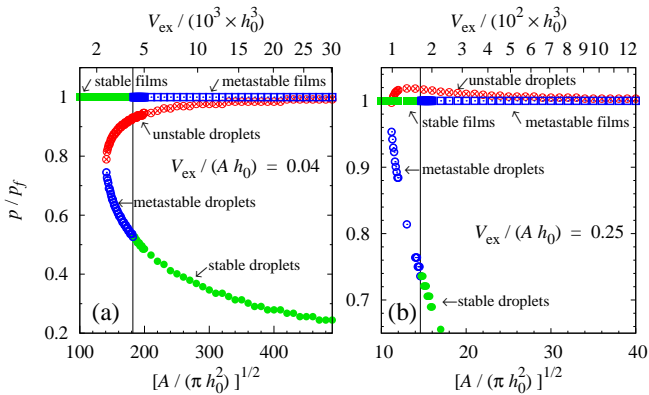}
\caption{
The pressure $p=-\Pi(\hpweq)$ (see Eq.~\eqref{elg_pw2})
(in units of $p_f=-\Pi(h_f)$)
calculated from the approximate free energy in Eq.~\eqref{conv_F} as
a function of the 
substrate size $A$ and of the excess volume $\Vex$ for (a) 
$\Vex/(A\,h_0)=0.04$ and (b)
$\Vex/(A\,h_0)=0.25$, i.e., for a fixed homogeneous film
thickness $h_f=1.04\,h_0<h_i$ and $h_f=1.25\,h_0>h_i$,
respectively, and $\phi_0/\sigma =0.5$\/. Note that for a fixed
ratio $\Vex/(A\,h_0)=v_\text{ex}$ one has
$\Vex/h_0^3=\pi\,v_\text{ex}\,x^2$ with
$x=[A/(\pi\,h_0^2)]^{1/2}$\/. 
Open blue and full green symbols indicate metastable and stable
states, respectively, and circles with crosses indicate unstable droplet
solutions. Stable and metastable flat film
solutions are indicated by boxes and stable and metastable droplets by
circles.
The vertical line indicates the morphological transition between
stable films and stable droplets as obtained via numerical
comparison of the corresponding two free energies.
(A Maxwell construction for
determining this transition point is
discussed in, cf., Fig.~\ref{maxwellfig}\/.)
$A \to\infty$ corresponds to the macroscopic limit\/.
(a) For $h_f<h_i$, as in the present case corresponding to
$\Vex/(A\,h_0)=0.04$, the flat
film solution is stable or metastable for all
$A$\/. Droplets (lowest branch) occur for
$\sqrt{A/(\pi\,h_0^2)}\gtrsim 140$ and they are stable for
$\sqrt{A/(\pi\,h_0^2)}\gtrsim 180$\/. 
(b) Within the present free energy approximation, the flat film solution is
stable or metastable (although it should be unstable according to
Eq.~\eqref{crit1}) and there is an unphysical unstable branch of
droplet solutions (top branch)\/. Droplets are stable or metastable
for for all substrate sizes $A$\/.
\label{fig_dhf-laplace}
}
\end{figure}

In Fig.~\ref{fig_Fmacro} the excess volume is chosen such that
$h_f<h_i$, i.e., according to Eq.~\eqref{crit1} the film
configuration is \textit{linearly} stable. Nonetheless, the droplet
solution is the global minimum of $F(\alpha,h_w)$\/. However, as shown in
Fig.~\ref{fig_dhf-laplace}(a) there is a minimal
droplet size $\Vex$ below which droplets cannot exist: reducing the
droplet size the Laplace pressure in the droplet increases until it
cannot be counterbalanced by the negative disjoining pressure
$-p=\Pi(\hpw^{eq})$ (see Eq.~\eqref{elg_pw2}) in
the film ($\Pi(z)$ has a minimum of \textit{finite} depth (see
Fig.~\ref{fig_eip_model})) and the droplet drains into
the film. In Fig.~\ref{fig_dhf-laplace}(a) there is also a second
branch of droplet solutions which are unstable and which have a
pressure $p$ intermediate between the pressure
of the metastable or stable droplets and of the flat film.
For a given value of $\Vex$ such a droplet
solution corresponds to the
saddle point in the two-dimensional parameter space between the two
(local) minima given by the droplet solution and the flat film
solution. Upon reaching the
macroscopic limit, the unstable droplet branch asymptotically
approaches the flat film pressure from below
(Fig.~\ref{fig_dhf-laplace}(a))\/. This means, that the
thickness $\hpweq$ of the wetting film surrounding the unstable droplets
approaches the thickness $h_f = \Vex/A + h_0$ of the flat film
solution. Therefore the volume inside the
unstable droplets (i.e., above $\hpw$) decreases monotonically as
the macroscopic limit is approached.
Figure~\ref{fig_dhf-laplace} corresponds to Fig.~12 in
Ref.~\cite{dutka12} where, however, the
volume rather than the
pressure is plotted as a function of the substrate size
without discussing the stability of the solutions.
We conclude that in this respect in essence
there is no qualitative difference between
the quasi two-dimensional ridges studied in
Ref.~\cite{dutka12} and the three-dimensional systems studied here.

For large excess volumes with $h_f>h_i$, according to
Eq.~\eqref{crit1}, the flat film solution is
linearly unstable.
Therefore it should represent a saddle point or a maximum in the
free energy landscape. The
droplet solution should represent the global minimum. However, as
shown in Fig.~\ref{fig_dhf-laplace}(b) the flat film solution for
$h_f=1.25\,h_0>h_i$ is either stable or metastable, but not
unstable within this only two-dimensonal parameter space
considered here. In addition there is an unphysical branch of
unstable droplet solutions with pressures above the pressure of the
flat film solution. The reason for this artefact is, that a slightly
undulated film cannot be represented in this two-dimensional
parameter space; but spinodal
dewetting occurs via the growth of such small perturbations. According
to Eq.~\eqref{crit2}, the critical substrate size, below which the
instability is suppressed by the finite size effects, is
$\sqrt{A/(\pi\,h_0^2)}\approx 6$, i.e., much smaller than
$\sqrt{A/(\pi\,h_0^2)}\approx 11.2$, the smallest
substrate size for which the present
two-dimensional parameter space analysis predicts the
existence of droplet solutions (see
Fig.~\ref{fig_dhf-laplace}(b))\/. 
In view of this inconsistency we conclude that the results obtained
within this approximate scheme for very
small substrate sizes are unreliable. However, the actual stability
of droplets in the macroscopic limit is correctly covered within this model.

At the morphological transition a flat film and a droplet
of equal volume have the same free energy but different pressure.
In the theory of thermodynamic phase transitions, 
it is common to consider transitions between states of
different volume (or density) but equal pressure (or more
general, between states with equal intensive state variables
but distinct extensive ones)\/.
These states can spatially coexist with each other. However, the
morphological transition between a flat film and a droplet is of a
different nature in the sense that the droplet solution and the
flat film solution do not coexist with each other in space: 
the system as a whole switches from one
solution to the other. This is not to be confused
with the coexistence between a droplet and the wetting film to
which it is connected. While the pressure in the wetting film and
the pressure
in the droplet are equal, this droplet
configuration does not represent a bona fide
thermodynamic phase: its pressure changes with size
whereas from a
proper thermodynamic phase one would expect to be able to produce
systems of different size but with the same pressure. In fact,
Eq.~\eqref{hamilgk} has the structure of a Ginzburg-Landau
Hamiltonian but the potential $\Phi(h)$ has its second minimum at
$h\to\infty$\/. In this sense the droplet 
as a whole amounts to an interfacial region.

\begin{figure} 
\includegraphics[width=\linewidth]{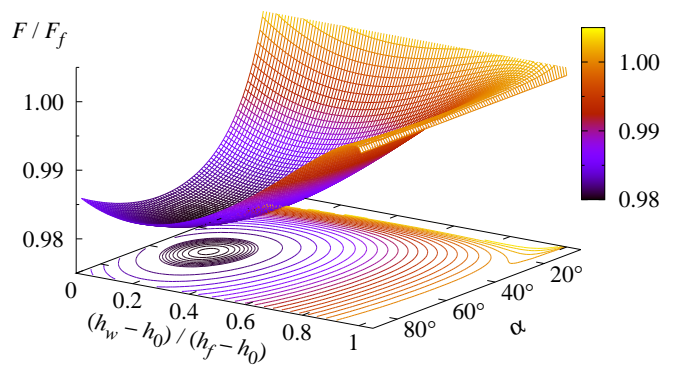}
\caption{
The approximate interfacial free energy $F(\alpha,\hpw)$ for
$A/(\pi h_0^2)=100^2$, $\Vex/(A\,h_0)=0.10$, and
$\phi_0/\sigma =0.5$.
$F_f=A\,(\sigma +\phi(h_f))$ is the free energy of
the flat film solution for these parameters; $h_f-h_0 = \Vex/A$\/. 
The global minimum
representing a stable nano-droplet is located at
$(\hpweq-h_0)/(h_f-h_0)\approx 0.2$ and
$\aleq\approx57^\circ$, i.e., 
close to but smaller than the macroscopic equilibrium contact angle
$\theq=60^\circ$ for this system. The flat film solution
$\hpweq=h_f$ (so that $(\hpweq-h_0)/(h_f-h_0)=1$) is
metastable; for this solution there is no dependence on $\alpha$\/.
Contour lines are shown in the range $0.9802$ to $0.9809$ in steps of $0.0001$ and
from $0.981$ to $1.005$ in steps of $0.001$\/. 
}
\label{fig_Fvex0.10}
\end{figure}

\begin{figure}
\includegraphics[width=\linewidth]{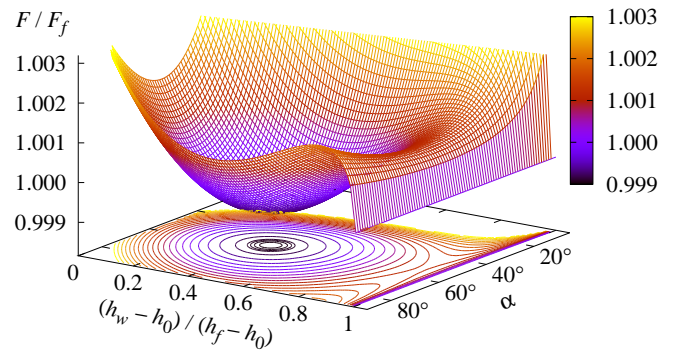}
\caption{
The approximate free energy $F(\alpha,\hpw)$ as defined in
Eq.~\eqref{conv_F} for $A/(\pi h_0^2)=100^2$,
$\Vex/(A\,h_0)=0.06$, and $\phi_0/\sigma =0.5$ (i.e., for the same parameters
as in Fig.~\ref{fig_Fvex0.10} but for a smaller value of $\Vex$).
$F_f=A\,(\sigma +\phi(h_f))$ is the free energy of
the flat film solution for these parameters; $h_f-h_0 = \Vex/A$\/.
The contact angle corresponding to the global minimum is
$\aleq\approx55^\circ$ at $(\hpweq-h_0)/(h_f-h_0)\approx
0.04$,
i.e., smaller than in Fig.~\ref{fig_Fvex0.10}\/. 
The flat film solution $\hpweq=h_f$ (so that $(\hpweq-h_0)/(h_f-h_0)=1$) 
is metastable and exhibits no dependence on $\alpha$\/.
Contour lines are shown in the range $0.99902$ to
$0.9991$ in steps of $0.00002$ and from $0.9992$ to $1.003$ in
steps of $0.0002$\/. 
}
\label{fig_Fvex0.06}
\end{figure}

\begin{figure}
\includegraphics[width=\linewidth]{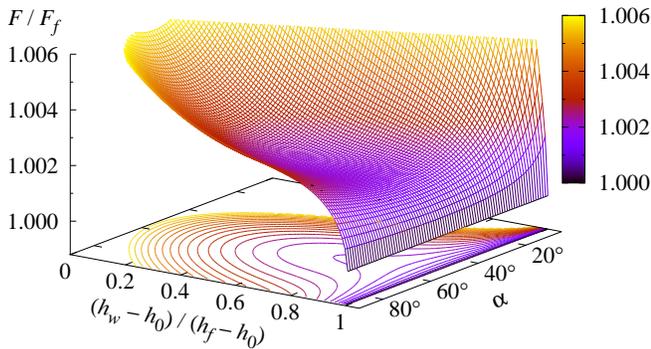}
\caption{
The approximate interfacial free energy $F(\alpha,\hpw)$ for
$A/(\pi h_0^2)=100^2$, $\Vex/(A\,h_0)=0.048$, and
$\phi_0/\sigma =0.5$ (i.e., for the same parameters
as in Figs.~\ref{fig_Fvex0.10} and \ref{fig_Fvex0.06} but for a
smaller value of $\Vex$).
$F_f=A\,(\sigma +\phi(h_f))$ is the free energy of
the flat film solution for these parameters; $h_f-h_0 = \Vex/A$\/. 
For this excess volume  the droplet solution has disappeared and the
flat film solution $\hpweq=h_f$ (so that $(\hpweq-h_0)/(h_f-h_0)=1$)
is the global minimum.  
Contour lines are shown in the range $0.999$ to $1.006$ 
in steps of $0.00025$\/. 
\label{fig_Fvex0.05}}
\end{figure}

The free energy landscape for finite systems with various excess
volume ratios $\Vex/(A\,h_0)$ are shown in
Figs.~\ref{fig_Fvex0.10}--\ref{fig_Fvex0.05}\/.
For the large excess volume in Fig.~\ref{fig_Fvex0.10}, the droplet
configuration with $\aleq\approx 57^\circ$ and $\hpweq-h_0\approx
0.2\,(h_f-h_0)$ is the global minimum. The flat film solution with
$h_f=1.1\,h_0<h_i$ is linearly stable as expected for the chosen
effective interface potential (see Eq.~\eqref{crit1})\/. Upon decreasing
the excess volume the
free energy of the droplet solution increases and the minimum
becomes shallower (see Fig.~\ref{fig_Fvex0.06})\/. At a certain
excess volume,
the flat film solution becomes the stable solution and the droplet
solution becomes metastable.
Reducing the excess
volume even further, the free energy minimum
corresponding to a droplet solution becomes more and more shallow
until it finally merges with the corresponding saddle point (see
Fig.~\ref{fig_Fvex0.05}), and vanishes completely. This leaves
the film solution as the only stable solution.

\begin{figure} 
\includegraphics[width=\linewidth]{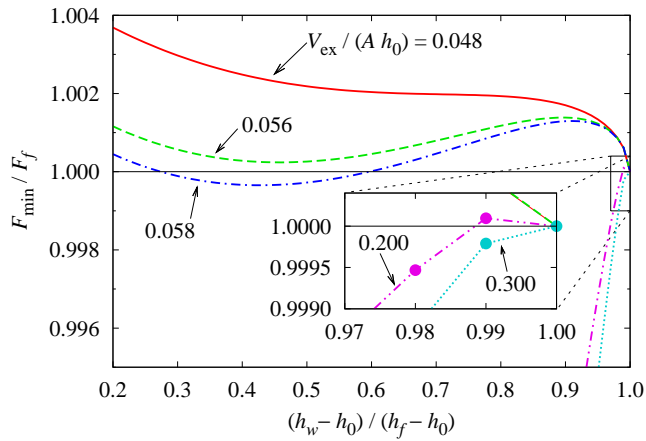}
\caption{
The minimal free energy $F_\text{min}(\hpw)=\min_{\alpha} F(\alpha,\hpw)$ as a
function of the wetting film thickness $\hpw$ for $A/(\pi
h_0^2)=100^2$, $\phi_0/\sigma =0.5$, and several values of
$\Vex/(A\,h_0)$ ranging from $0.048$ to $0.3$\/. 
$F_f=A\,(\sigma +\phi(h_f))$ is the free energy of
the corresponding flat film solution for these parameters. 
The morphological transition between flat films and nano-droplets
occurs between $\Vex/(A\,h_0)=0.058$ and $0.056$\/. 
For $\Vex/(A\,h_0)>0.2$ the flat film solution appears to becomes
unstable as expected from Eq.~\eqref{crit1} (in the inset see
the enlarged view of the region near $h_w=h_f$)\/. The symbols
indicate the points calculated numerically. For
$\Vex/(A\,h_0)=0.3$ a small barrier cannot
be ruled out on the basis of the available numerical data.
\label{fig_as100}
}
\end{figure}

This morphological transition is  visualized even
better by forming vertical cuts of
the free energy landscape 
at fixed $\hpw$, i.e., parallel to the $\alpha$-axis and by
seeking the minimum of the free energy $F$ within each cut as a
function of $\alpha$\/. This renders $F_{\mathrm{min}}(\hpw) =
\min_\alpha F(\alpha,\hpw)$\/.
In Fig.~\ref{fig_as100} the corresponding the minimal free energy 
$F_{\mathrm{min}}(\hpw)$ is shown as a 
function of the wetting film thickness $\hpw$\/. The energy scale
is normalized by the free energy $F_f(\Vex)$ of the corresponding
flat film solution (compare Figs.~\ref{fig_Fmacro} to
\ref{fig_Fvex0.05})\/. For
very small excess volumes the free energy as a function of the
wetting film thickness is monotonically decreasing and
the only minimum which occurs is the one
corresponding to a flat film of thickness $h_f$ so that
$(h_w-h_0)/(h_f-h_0)=1$\/. For intermediate
excess volumes ($0.05\lesssim \Vex/(A\,h_0)\lesssim 0.058$ in
Fig.~\ref{fig_as100}) there is a second minimum corresponding to a
metastable droplet. With increasing $\Vex$ this droplet minimum
deepens until it is as deep as the minimum
corresponding to the flat film (at $\Vex/(A\,h_0)\approx
0.057$)\/.
 This marks the point of the
morphological transition between a flat film and a droplet solution.
Increasing the excess volume even further the droplet solution
becomes more stable. According to the inset of Fig.~\ref{fig_as100}
it seems that the flat film solution (i.e., $(\hpw-h_0)/(h_f-h_0)=1$)
becomes unstable for $\Vex/(A\,h_0)\ge 0.2$ as expected from
Eq.~\eqref{crit1}\/. However, for this latter value a tiny free
energy barrier cannot be ruled out on the basis of the available numerical data.

\begin{figure} [t]
\includegraphics[width=\linewidth]{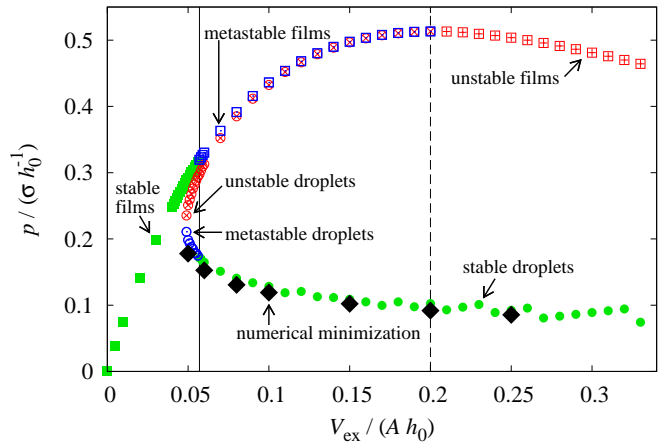}
\caption{
The pressure $p$ in units of $\sigma/h_0$ as a function of $\Vex/(A\,h_0)$ for a fixed
substrate size, $A/(\pi\,h_0^2)=100^2$, and
$\phi_0/\sigma =0.5$ as obtained from the approximate free energy 
expression $F(\alpha,\hpw)$\/.
Global minima (full green), local minima (open blue), and saddle
points or local maxima (symbols with red crosses) are shown.
The upper branch corresponds to flat films (boxes)
and the lower one to nano-droplets (circles)\/. Within the
reduced model we have $p=-\Pi(\hpw^{eq})$ (see
Eq.~\eqref{elg_pw2})\/.
The pressure values obtained from a numerical minimization of
Eq.~\eqref{hamil} (black diamonds; $p\approx-\Pi(\hpw^{eq})$, see
Eq.~\eqref{elg_pw}) agree well with the results obtained from 
the approximate free energy.
The dashed vertical line indicates the volume
at which $h_f=h_i$\/. At this volume the unstable droplet branch
merges with the flat film branch. The full vertical line indicates 
the morphological transition between the film and the 
droplet configurations.
\label{fig_as100-laplace}
}
\end{figure}

\begin{figure}
\includegraphics[width=\linewidth]{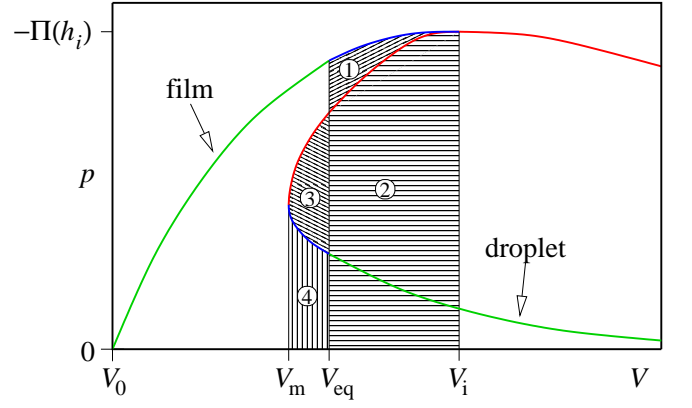}
\caption{\label{maxwellfig} Sketch of the Maxwell construction
leading to the position of the thin full vertical line in
Fig.~\ref{fig_as100-laplace}\/. The color code corresponds to the
one in Fig.~\ref{fig_as100-laplace}: green,
blue, and red indicate stable,
metastable, and unstable states, respectively.
$V_0$ denotes the volume of a film
of thickness $h_0$, $V_m$ is the minimal volume required to form a
droplet, $V_\text{eq}$ is the volume at which the free
energies of the flat film and of the stable droplet are equal.
For $V\nearrow V_i$ the branch of metastable flat films turns into
a branch of unstable flat films. There also the branch of
unstable droplets merges into the flat film branch.
}
\end{figure}

Figure~\ref{fig_as100-laplace} shows the pressure $ p$ as a
function of the excess volume for a homogeneous film of thickness
$h_f$ (upper curve) and for the droplet solution (lower curve)\/.
The upper branch is exact while the lower branch is calculated 
by minimizing the approximate expression for the free energy 
$F(\alpha,\hpw)$ defined in Eq.~\eqref{conv_F}\/.
According to Eq.~\eqref{elg_pw2}, for both
branches one has $p=-\Pi(\hpw^{eq})$\/.
Figure~\ref{fig_as100-laplace} also shows pressure values obtained by 
numerical minimization of the
full functional $\maF$ (for which, according to
Eq.~\eqref{elg_pw}, $p\approx-\Pi(\hpw^{eq})$)\/. 
The pressure in the flat films (upper curve)
is given by $p=-\Pi(h_f)$ with $h_f/h_0=\Vex/A+1$ (see
Eq.~\eqref{elg_pw2}) and it has a maximum at
$\Vex/(A\,h_0)=0.2$, corresponding to $h_f=h_i$\/. For excess
volumes smaller than $\Vex/(A\,h_0)=0.2$ the flat film solution
is metastable or stable. For larger excess volumes, the flat film solution is
linearly unstable. However, the spinodal wavelength is extremely
large close to the pressure maximum such that, according to Eq.~\eqref{crit2} 
and for the given substrate size, the instability actually sets in only for
$\Vex/(A\,h_0)>0.2011$\/. In Fig.~\ref{fig_as100-laplace} for $0.05<\Vex/(A\,h_0)<0.2$ there are two
curves below the curve corresponding to the flat film solution; the
upper one (red circles with crosses) corresponds to a saddle point
in the free energy landscape and the
lower one corresponds to a (potentially local) minimum.
Both branches represent droplet solutions. The
unstable branch ends at $\Vex/(A\,h_0)=0.2$, i.e., at the
maximum of the pressure in the flat film solution. The
three curves in Fig.~\ref{fig_as100-laplace}
form a hysteresis loop. The value of
$\Vex/(A\,h_0)$, at which the transition (thin vertical line in
Fig.~\ref{fig_as100-laplace})
between a flat film and a droplet occurs, can be obtained
either by comparing free energies directly or via a
Maxwell construction (see Fig.~\ref{maxwellfig})\/. The
latter can
be shown by integrating $p=\partial \maF/\partial V = \partial
\maF/\partial \Vex$ (due to Eq.~\eqref{hamil} and since
$dV/d\Vex =1$ due to Eq.~\eqref{defVex}) along
$p(V)$: $\mathcal{F}(V)-\mathcal{F}(V_0) = \int_{V_0}^{V}
p(V)\,dV$\/. Starting the integration at the volume $V_\text{eq}$ at which the free
energy of the film (upper branch) and the stable droplet (lowest
branch) are equal (see Fig.~\ref{maxwellfig}) one integrates up to $V=V_i$, i.e., the volume of a
film of thickness $h_i$ at which the unstable droplet branch merges
with the film branch. The result, i.e., the sum of area (1) and (2) in
Fig.~\ref{maxwellfig}, is the difference of the free energies of a
film with volume $V_i$ and a film with volume $V_{\text{eq}}$\/.
At $V_i$ one switches to the unstable
droplet branch and integrates down
to its end at $V_m$\/. The result is the difference between area
(1) and the sum of area (3) and area (4)\/. From
there one continues on the
metastable droplet branch up to $V_{\text{eq}}$, which adds area
(4)\/. As a result, the difference of the free energy of a flat film
of volume $V_\text{eq}$ and a stable droplet of the same volume is
the difference between area (1) and area (3)\/.
For the chosen model interface
potential in Eq.~\eqref{eip_model} the flat film solution becomes
linearly unstable at the value of $\Vex/(A\,h_0)$
(i.e., $0.2$ in Fig.~\ref{fig_as100-laplace}), where the
unstable droplet curve merges with the flat film curve. 

\begin{figure}
\includegraphics[width=\linewidth]{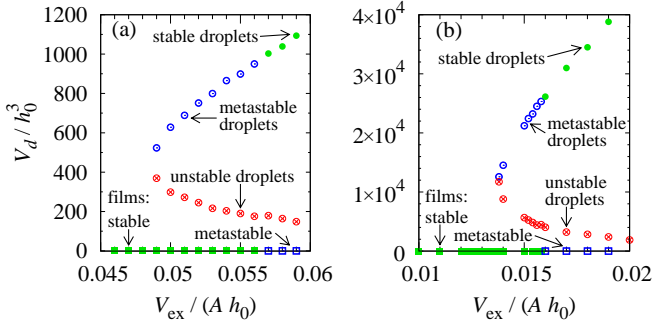}
\caption{The droplet volume $V_d=V-A\,\hpw$ as function of
$\Vex/(A\,h_0)$ (i.e., as a function of $V=\Vex+A\,h_0$)
for $\phi_0/\sigma
= 0.5$ and (a) $A/(h_0^2\,\pi) = 100^2$ and (b) $A/(h_0^2\,\pi) =
1000^2$ as obtained from the approximate expression
$F(\alpha,\hpw)$ for the free energy: for fixed
$V=\Vex+A\,h_0$ the free energy landscapes (see, e.g.,
Figs.~\ref{fig_Fvex0.10}--\ref{fig_Fvex0.05}) have
been calculated and the wetting film thicknesses
$\hpw$ of the droplet solutions---if they exist---have
been determined. The upper
branch corresponds to the stable (green) or metastable (blue)
droplet solution. For (a) this is the lower branch in
Fig.~\ref{fig_as100-laplace}\/. The points on the abscissa
correspond to stable (green) or metastable (blue) flat film
solutions. The comparison between (a) and (b) shows that the
minimal droplet size $V_d^c$ (for which the unstable and the
metastable droplet branches meet) increases upon
increasing the substrate area
while the corresponding excess volume $\Vex^c$ in units of the substrate
area decreases.
\label{fig_Vd}}
\end{figure}

In Fig.~\ref{fig_as100-laplace} the excess volume is expressed in
terms of 
the substrate area. In order to discuss whether the minimal droplet
size is determined by the interface potential or by the
substrate size, one could fix the excess volume $\Vex$ (as a
measure for the droplet size) and the
substrate potential and vary the substrate size $A$\/. But the excess
volume is defined as the fluid volume above the height $h_0$ (see
Eq.~\eqref{defVex}) and
increasing the substrate area $A$ for fixed $\Vex$ means effectively
reducing the droplet size.
The droplet volume $V_d=V-A\,\hpw$ above the height of the wetting
film $\hpw$ is a more suitable measure for the droplet size. For this reason 
in Fig.~\ref{fig_Vd} we
plot the droplet volume $V_d$ as a function of the excess volume
$\Vex$ for two substrate sizes. The data are obtained in the
following way: for each fixed value of $A$ and of $\Vex$
(i.e., for fixed total volume $V=\Vex+A\,h_0$) the interfacial
free energies as shown in
Figs.~\ref{fig_Fvex0.10}--\ref{fig_Fvex0.05} are calculated. The
position of local and global minima and of saddle
points (corresponding to stable, metastable, and unstable droplet or flat film
solutions) are determined numerically, in particular the wetting film
thickness
$\hpw$ from which one can determine the droplet volume 
$V_d=\Vex-A\,(\hpw+h_0)$\/. As in
Fig.~\ref{fig_as100-laplace}, for large $\Vex$ there are 
three branches of solutions (flat film solutions with $V_d =0$, 
unstable droplet solutions, and metastable or stable droplet
solutions). For small $\Vex$ there are only flat film solutions.
The size $V_d^c=V_d(\Vex=\Vex^c)$ of the smallest metastable
droplet (which is identical to the size of the largest unstable
droplet) increases
with the substrate area, as well as the 
value $\Vex^c$ of the corresponding excess
volume. However, $\Vex^c/(A\,h_0)$ decreases
upon increasing $A$ (compare
Figs.~\ref{fig_Vd}(a) and (b))\/. This means, that the thickness
$h_f^c = \Vex^c/A$ 
of the flat film solution corresponding to the minimal droplet also
decreases upon an increase of the substrate area.

\begin{figure}
\includegraphics[width=\linewidth]{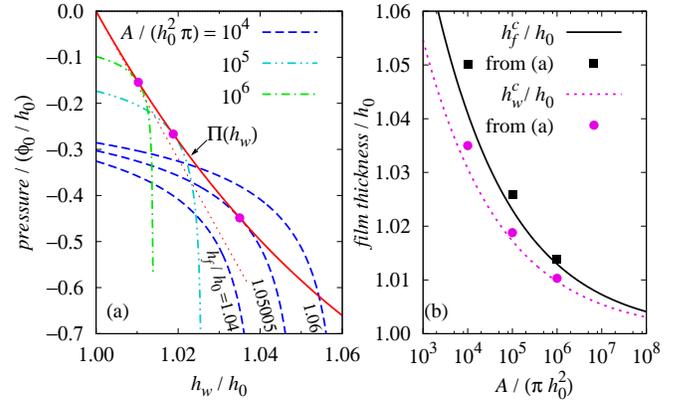}
\caption{\label{lowercrit} (a) Disjoining pressure $\Pi(\hpw)$
with $h_0<\hpw <h_f$ (calculated for the
model potential in Eq.~\eqref{eip_model}) in the wetting film
(solid red line) and the right hand side of Eq.~\eqref{pressure}
with $R(\hpw,h_f,A)$ from Eq.~\eqref{volume} (dashed blue
lines) 
for $A/(\pi\,h_0^2) = 10^4$ and $\phi_0/\sigma = 0.5$ as a function
of the wetting film thickness $\hpw$ for $h_f=1.04\,h_0$,
$h_f=h_f^c=1.05005\,h_0$, and
$h_f=1.06\,h_0$, which fixes $V=A\,h_f$ for a given $A$ (see
Eq.~\eqref{defVex})\/. We also show the right hand side of
Eq.~\eqref{pressure} for $A/(\pi\,h_0^2) = 10^5$ and
$h_f=h_f^c=1.02587\,h_0$ (dash-double-dotted cyan line), as well as
for $A/(\pi\,h_0^2) = 10^6$ and 
$h_f=h_f^c=1.01394\,h_0$ (dash-dotted green line)\/.
The thin dotted red line shows the linear fit to
$\Pi(\hpw)$ at $h_w=h_0$\/. 
For $h_f=h_f^c(A)$ the curves $\Pi(\hpw)$ and the one for
the right hand side of  Eq.~\eqref{pressure} touch each other at a
single point at $h_w=h_w^c(A)$ indicated by a magenta
circle. 
(b) $h_f^c$ and $\hpw^c$ as a function of the substrate area
$A$ in units of $\pi\,h_0^2$ as obtained graphically from (a) (black squares and magenta circles,
respectively) and from the analytic approximation described in the
main text (full black and dotted magenta line, respectively)\/.
}
\end{figure}

The
nonexistence of droplet solutions for too small values of $\Vex$ can be
rationalized by considering a further simplified 
reduced expression for the free energy.
Neglecting the influence of the disjoining pressure on the
spherical cap the minimization problem for
$F(\alpha,\hpw)$ yields 
(see Eq.~\eqref{conv_F} and up to the constant
substrate-liquid surface tension) 
\begin{equation}
F = (A-r^2\,\pi)\,\left[\sigma +\phi(\hpw)\right]+ \sigma\,S_d,
\end{equation}
with $r=\sqrt{2\,h_d\,R-h_d^2}$ denoting the base radius of the
\textit{d}rop
(taken at $z=\hpw$) and $S_d = 2\,\pi\,R\,h_d$ denoting the
surface area of a spherical cap of height $h_d$ and radius $R$\/. The
volume of the spherical cap is given by $V_d =
\frac{\pi}{3}\,h_d^2\,(3\,R-h_d)$ and the total fluid volume by $V=V_d
+ A\,\hpw$\/. It is convenient to write the volume constrained free
energy 
\begin{multline}
F(h_d,\hpw) =
\left[A\,\left(1+\frac{2\,\hpw}{h_d}\right)-\frac{2\,V}{h_d}+\frac{\pi}{3}\,h_d^2\right]\,\phi(\hpw)\\
+\sigma\,\left(A+\pi\,h_d^2\right)
\end{multline}
as a function of the droplet height $h_d$ rather
than the droplet contact angle $\alpha$\/. The minimum of $F(h_d,\hpw)$ 
follows from the zeroes of its first derivatives with respect to
$h_d$ and $\hpw$\/. Using the above expressions for $V$ and
$V_d$ one obtains from $\partial F(h_d,\hpw)/\partial h_d =0$ 
\begin{equation}
\label{elgeqH}
R\,\phi(\hpw)+\sigma\,h_d = 0.
\end{equation}
Using this expression together with the above expressions for $V$ and
$V_d$ one obtains from $\partial
F(h_d,\hpw)/\partial \hpw=0$, after
reintroducing $\alpha$ via the geometric condition
$r=R\,\sin\alpha$,
\begin{equation}
\Pi(\hpw)=
 -\frac{2\,\sigma}{R\,\left(1-\frac{\pi\,R^2\,\sin^2\alpha}{A}\right)} .
\label{pressure}
\end{equation}
Apart from a
small correction (which is small if $A$ is large
compared with the base area
$\pi\,R^2\,\sin^2\alpha$ of the droplet) Eq.~\eqref{pressure}
tells that the
disjoining pressure in the film and the Laplace
pressure $2\,\sigma\,H_h = -2\,\sigma/R$ (see
Eq.~\eqref{eqcurvature}) in the
droplet are equal (according to Eq.~\eqref{elg} both are equal
to $p$)\/. 
Using the geometric relation $\cos\alpha=1-h_d/R$ in
Eq.~\eqref{elgeqH} we also get 
\begin{equation}
\label{cosalpha}
\cos \alpha = 1+\frac{\phi(\hpw)}{\sigma}.
\end{equation}
In the macroscopic limit $R\to\infty$ in Eq.~\eqref{pressure}
implies $\Pi(\hpw)\to 0$, i.e., $\hpw\to h_0$ so that
$\phi(\hpw)\to \phi(h_0)=-\phi_0$ (see
Fig.~\ref{fig_eip_model}), and therefore
$\alpha\to \theq$ (see Eq.~\eqref{young})\/.
As a function of $\alpha$, $R$, and $\hpw$
the total conserved fluid volume is
\begin{equation}
V= A\,\hpw + \frac{\pi\,R^3}{3}\,(2 +
\cos\alpha)\,(1-\cos\alpha)^2.
\label{volume}
\end{equation}
For a given value of $\alpha$ Eqs.~\eqref{pressure} and
\eqref{volume} provide solutions for
$\hpw$ and $R$ only if $V$ is
sufficiently large. The thickness
$\hpw$ can only vary between $h_0$ (i.e., the whole excess volume is
concentrated in the droplet) and $h_f=V/A$ (i.e., there is no droplet)\/. For
$\hpw=h_0$ the disjoining pressure $\Pi$ in the film is zero while the
Laplace pressure $2\,\sigma\,H_h=-2\,\sigma/R$ in the droplet is negative. Both
become more negative
for increasing $\hpw$ because the droplet shrinks and
$\Pi'(h_0)<0$\/. The
Laplace pressure diverges to $-\infty$ as $\hpw\to h_f=V/A$ because the
droplet volume $V_d$ (and therefore the droplet radius $R$)
vanishes
in this limit and $H_h=-1/R$\/. But the
disjoining pressure is bound from below. 
With $\arccos[1+\phi(\hpw)/\sigma]$ (see
Eq.~\eqref{cosalpha})
Eq.~\eqref{volume} can be solved
for $R$ yielding 
$R(\hpw,V,A)$ or 
\begin{equation}
R(\hpw,h_f,A) = \sqrt[3]{
\frac{3\,A\,(h_f-\hpw)}{\pi\,\left[3+\frac{\phi(\hpw)}{\sigma}\right]\,
\left[\frac{\phi(\hpw)}{\sigma}\right]^2}
}
\end{equation}
due to $V=A\,h_f$\/.
Accordingly, one can consider both sides of
Eq.~\eqref{pressure} as a function of $\hpw$ as shown in
Fig.~\ref{lowercrit}(a) where $\alpha$ is approximated
by $\theq=\arccos(1-\phi_0/\sigma)$\/.
The right hand side of Eq.~\eqref{pressure} increases (decreases in
absolute value) upon increasing
$h_f/h_0=V/(A\,h_0)$\/. 
For large $A/R^2$ and
$h_w\approx h_0$, the right hand side of Eq.~\eqref{pressure} is
approximately given by
\begin{equation}
-\frac{\phi_0}{h_0}\,\frac{2}{\sqrt[3]{3}}\,\sqrt[3]{
\frac{\frac{\sigma}{\phi_0}\,\left(3+\frac{\phi_0}{\sigma}\right)}
{\frac{A}{\pi\,h_0^2}\,\frac{h_f-\hpw}{h_0}}}.
\end{equation}
The two curves only intersect
if the fluid volume (or $h_f = V/A$) is sufficiently large
(see the three blue dashed curves in Fig.~\ref{lowercrit}(a))\/. 

For sufficiently large excess volumes, i.e., for sufficiently
large $h_f$ there are two intersections in
Fig.~\ref{lowercrit}(a)\/. Because for fixed total volume $V$ increasing 
$h_w$ (i.e., increasing the amount of liquid in the film) means
decreasing the droplet volume, the intersection at the larger values of
$h_w$ corresponds to the unstable solution while the intersection
at the smaller value of $\hpw$ corresponds to the stable droplet
solution. (The unstable droplet is always smaller than the stable
one.)
In the macroscopic limit $A/h_0^2\to\infty$ with fixed
$h_f=h_0+\Vex/A$ (see Eq.~\eqref{macro_limit}) the stable
solution moves to $h_w \to h_0$ \/.
This means that the volume of the stable droplet gets very
large because due to $\hpw\to h_0$ the whole excess volume goes
into the droplet.
In the macroscopic limit, the unstable solution moves to $h_w \to
h_f$\/. We can obtain the corresponding leading 
behavior by the following
procedure. First we insert $R(h_w,V,A)$ as obtained from
Eq.~\eqref{volume} into Eq.~\eqref{pressure} and we replace
$\cos\alpha$ by the expression in Eq.~\eqref{cosalpha}\/. After
substituting $V = A\,h_f$ we expand both sides in powers of
$\hpw-h_f$ and we obtain in leading order $h_f-\hpw \sim
1/A$ \footnote{This calculation can be significantly simplified
by approximating $\alpha \approx \theq$ (which is independent of
$\hpw$) and by neglecting the term $\sim A^{-1}$ in the denominator
of the right hand side of Eq.~\protect\eqref{pressure}\/.}\/. As a
consequence, in the macroscopic limit the volume
$V_d=A\,(h_f-\hpw)$ of the unstable droplet should
converge to a finite value. However, 
this primitive model only applies to large
droplet volumes and therefore this result for unstable drops 
might turn out to be an artefact of the approximations used.

As shown in Fig.~\ref{lowercrit}(b) the critical
average film thickness $h_f^c = V^c/A$ required for forming a droplet
decreases as a function of the substrate area. For very large $A$
both $h_f^c$ and the corresponding wetting film thickness $\hpw^c$
corresponding to the
smallest possible droplet are very close to $h_0$ such that
in Eq.~\eqref{pressure} one can expand $\Pi(\hpw)$
around $h_0$
(see the thin dotted line in Fig.~\ref{lowercrit}(a))\/. 
If one makes the additional approximations of using $\alpha\approx \theq$
and of reducing the right hand
side of Eq.~\eqref{pressure} to the Laplace pressure by neglecting the term $\sim A^{-1}$ in the denominator
of the right hand side of Eq.~\protect\eqref{pressure}, 
one can determine $h_f^c$ and $\hpw^c$ analytically with the result 
$h_{f/w}^c -h_0 \propto A^{-1/4}$ (see
Fig.~\ref{lowercrit}(b))\/. The drop volume is $V_d = V-A\,\hpw =
A\,(h_f-\hpw)$\/. Thus the volume of the smallest
possible droplet diverges for $A\to\infty$ as
$V_d^c=(h_f^c-\hpw^c)\,A \propto A^{3/4}$\/.

\section{Summary and conclusions}

We have studied the stability of nonvolatile flat films
and droplets
on smooth and chemically homogeneous substrates with
finite surface area $A$\/.  The analysis is based on density functional
theory within the so-called
sharp kink approximation, i.e., by minimizing the effective
local
interface Hamiltonian with the effective interface potential
shown in Fig.~\ref{fig_eip_model}\/.

The stability of flat films and of
nano-droplets is strongly affected by finite size effects. We have
shown that in these systems
(i) spinodal dewetting can occur only if the substrate area $A$ is large
enough to support the shortest unstable wavelength,
(ii) there is a minimal size for droplets connected to a
surrounding wetting
layer,
(iii) droplets are unstable with respect to drainage into a
connected wetting films if the substrate area is too large,
and (iv) that fluctuations of the droplet shape 
under the constraint of a fixed volume
are more likely than volume fluctuations.

Our findings are manifestations of the general rule that
long-wavelength instabilities are suppressed by finite size
effects. The shortest instability wavelength
$L_c=2\,\pi\,\sqrt{\sigma/|\phi''(h_f)|}$ of
spinodal dewetting depends on the material properties, i.e., on the
surface tension $\sigma$ and on the effective interface potential
$\phi(z)$, as well as on the average film thickness $h_f=V/A$,
whereas $V$ is the conserved total liquid volume. In particular
for film thicknesses close to inflection points of $\phi(z)$ and for
thick films this wavelength becomes very large. For
differentiable effective interface potentials the second derivative
has a maximum (typically at a thickness of a few $h_0$ where
$\phi'(h_0)=0$)\/. Therefore the
spinodal wavelength $L_c$ of films with the corresponding thickness
has a minimum. Experimentally spinodal
wavelengths of the order of microns have been reported
\cite{seemann01b,seemann01d}\/. This means that spinodal dewetting can
be suppressed by structuring the surface, e.g.,
by a periodic pattern of hydrophilic and hydrophobic stripes,
the latter ones with a
width smaller than $L_c$\/. The width of the hydrophilic
stripes which is necessary to stabilize the film has to be
determined separately.

We have calculated the shape of nano-droplets
numerically as shown in Fig.~\ref{fig_nanodrops} and we
have determined the thickness
$\hpw$ of the wetting film on which the nano-droplet
resides (see Table~\ref{table})\/. Using a subset of trial
function for the droplet shape which are parameterized by the
contact angle of the droplet and by the wetting film thickness
$\hpw$ (see Fig.~\ref{fig_profiles}) we have mapped the free
energy landscape of the
system (see Figs.~\ref{fig_Fmacro} and
\ref{fig_Fvex0.10}--\ref{fig_as100})\/.

In contrast to macroscopic drops (see Fig.~\ref{fig_Fmacro}), 
for nano-droplets the influence of
the wetting film to which they are connected cannot be neglected.
If the excess volume $\Vex=V-A\,h_0=(h_f-h_0)\,A$
is fixed, there is a minimal substrate
size below which no droplet solutions exist (see
Fig.~\ref{fig_dhf-laplace})\/.
Conversely, for a fixed substrate size $A$ one can find droplet solutions only
above a critical (excess) volume (see Figs.~\ref{fig_Fvex0.06} and
\ref{fig_as100-laplace})\/. This is reminiscent of classical
nucleation theory which also leads to the notion of a critical nucleus size.
However, in the latter case one usually considers unbounded systems such that
one cannot obtain stable droplet solutions at all. In the present
case, the conserved total volume of
fluid is distributed between a finite sized wetting film and a
droplet; this allows for stable droplet solutions. 

As illustrated in Fig.~\ref{maxwellfig} 
the volume $V$ (or excess volume $\Vex=V-A\,h_0$) at which the
free energy of the flat film solution (a film of homogeneous
thickness $h_f=V/A$) equals the free energy of the stable droplet
(indicated by a thin vertical line in
Fig.~\ref{fig_as100-laplace})
can be determined by a Maxwell construction. This construction is
based on the observation that the Lagrange multiplier $p$ (i.e.,
the pressure difference between the liquid and the vapor phase) is
given by $p=\partial \mathcal{F}/\partial V$, i.e., by the partial
derivative with respect to the chosen total volume $V$ (see
Eq.~\eqref{hamil})\/.

The size of the smallest possible
droplet increases (see Fig.~\ref{fig_Vd})
and the thickness of the wetting film surrounding the droplet
decreases upon increasing the substrate area (see Fig.~\ref{lowercrit})\/.
Within a suitable approximation of the free energy we have found
that the volume $V_d^c$ of the
smallest possible droplet diverges upon increasing the substrate
size $A$ as
$V_d^c/h_0^3 \propto (A/h_0^2)^{3/4}$\/. The proportionality
factor depends on the equilibrium contact angle $\theq$ and for
nonzero contact angles it is of the order of unity with
$0.077$ as a lower bound (realized at $\theq = 180^\circ$)\/.
For $h_0\approx 1\,\text{\AA}$ this means that the
minimal droplet volume on substrates of size
$A=1\,\mathrm{mm}^2$,  $1\,\mu\mathrm{m}^2$, and
$(100\,\mathrm{nm})^2$ equals that of a cube of edge length
$300\,\mathrm{nm}$, $10\,\mathrm{nm}$, and $1.4\,\mathrm{nm}$,
respectively. On the same substrate the volumes of the connected
wetting films of thickness
$1$~{\AA} fit into cubes of an
edge length of $4.6\,\mu\mathrm{m}$, $46\,\mathrm{nm}$, and
$10\,\mathrm{nm}$, respectively, i.e., they are much larger.
Our results show that nonetheless the finite extent of 
the substrate surface plays a significant role for the droplet
formation and the associated morphological phase transition.


\end{document}